\documentclass{l4dc2024}

\usepackage{times}
\usepackage{bm}
\usepackage[capitalize]{cleveref}
\crefname{equation}{}{}
\crefname{figure}{Figure}{Figures}
\usepackage{acronym}
\newcommand{\ddt}{\ensuremath{\frac{d}{dt}}}
\newcommand{\fupdate}{\ensuremath{\bm{f}}}
\newcommand{\fupdateof}[1]{\ensuremath{\fupdate\left(#1\right)}}

\newcommand{\xstate}{\ensuremath{\bm{x}}}
\newcommand{\xstatehat}{\ensuremath{\hat{\bm{x}}}}

\newcommand{\rkstage}[1]{\ensuremath{\bm{k}^{(#1)}}}
\newcommand{\rkstagehat}[1]{\ensuremath{\hat{\bm{k}}^{(#1)}}}

\newcommand{\rkstagedefect}[1]{\ensuremath{\bm{\delta}^{(#1)}}}
\newcommand{\rkresidual}[1]{\ensuremath{\bm{\zeta}^{(#1)}}}
\newcommand{\rkstagedefectall}{\ensuremath{\bm{\delta}}}
\newcommand{\rkresidualall}{\ensuremath{\bm{\zeta}}}

\newcommand{\errorcollocation}[1]{\ensuremath{\bm{\eta}^{(#1)}}}
\newcommand{\nstages}{\ensuremath{s}}

\newcommand{\jacobianfx}[1]{\ensuremath{\left.\frac{\partial \fupdate}{\partial \xstate}\right|_{#1}}}

\newcommand{\Rdim}[1]{\ensuremath{\mathbb{R}^{#1}}}

\newcommand{\norm}[2]{\ensuremath{\left\lVert#1\right\rVert_{#2}}}
\DeclareMathOperator*{\NN}{NN}

\title[Error estimation for PINNs with IRK methods]{Error estimation for physics-informed neural networks with implicit Runge-Kutta methods}

\coltauthor{%
 \Name{Jochen Stiasny} \Email{jbest@dtu.dk}\AND \Name{Spyros Chatzivasileiadis} \Email{spchatz@dtu.dk}\\
 \addr Department of Wind and Energy Systems\\
  Technical University of Denmark\\
  Kgs. Lyngby, DK-2800
}

\begin{document}

\acrodef{PINN}[PINN]{Physics-Informed Neural Network}
\acrodef{NN}[NN]{Neural Network}
\acrodef{ODE}[ODE]{Ordinary Differential Equation}
\acrodef{RK}[RK]{Runge-Kutta}


\maketitle

\begin{abstract}
  The ability to accurately approximate trajectories of dynamical systems enables their analysis, prediction, and control. Neural network (NN)-based approximations have attracted significant interest due to fast evaluation with good accuracy over long integration time steps. In contrast to established numerical approximation schemes such as Runge-Kutta methods, the estimation of the error of the NN-based approximations proves to be difficult. In this work, we propose to use the NN's predictions in a high-order implicit Runge-Kutta (IRK) method. The residuals in the implicit system of equations can be related to the NN's prediction error, hence, we can provide an error estimate at several points along a trajectory. We find that this error estimate highly correlates with the NN's prediction error and that increasing the order of the IRK method improves this estimate. We demonstrate this estimation methodology for Physics-Informed Neural Network (PINNs) on the logistic equation as an illustrative example and then apply it to a four-state electric generator model that is regularly used in power system modelling.%
\end{abstract}


\section{Introduction}\label{sec:introduction}

Differential equations provide a powerful modelling tool to describe how systems change. They are ubiquitous in science and engineering and are used to formulate the behaviour of dynamical systems \citep{strogatz_nonlinear_2015}. While instantaneous changes are simple to evaluate, there, usually, exists no analytical solution to these differential equations. To predict future states of the system, we revert to numerical integration methods, see for example \cite{hairer_solving_1993,hairer_solving_1996,iserles_first_2008} for detailed treatment of the topic. A key question is the estimation of the approximation error of an integration scheme---both for a single time step and multiple time steps. For established schemes such as \ac{RK} methods, this estimation hinges around the local approximation error $\varepsilon$ which is closely related to the step size $h$. This relation is expressed as the order $p$ of a scheme $\varepsilon = \mathcal{O}\left(h^{p+1}\right)$ and it describes the error characteristic when $h$ is sufficiently small \citep{hairer_solving_1993}. To obtain absolute error estimates, an additional \ac{RK} scheme can be constructed to form an embedded \ac{RK} method \citep{hairer_solving_1993}. The resulting error estimates are crucial in choosing time step sizes that yield a balance between a sufficiently accurate approximation and the required computational cost. 



For dynamical systems found in power systems, the computational burden of solving the associated differential equations is significant---\cite{konstantelos_implementation_2017} provide an indication of the scales. Grid operators, who need to ensure system stability, rely heavily on these numerical integration methods to choose control actions, but the computational burden becomes a limiting factor. Therefore, the choice of the integration method and accurate error estimate are critical; this becomes even more relevant in light of the energy transition with additional dynamic phenomena \citep{hatziargyriou_definition_2021}. \ac{NN}-based approaches such as in \cite{cui_frequency_2023,moya_dae-pinn_2023,stiasny_pinnsim_2023} could have upsides in terms of prediction speed, however, the question of estimating and controlling the approximation error poses a major challenge to their adoption. Motivated by this example, this work shall provide a novel error estimation procedure to understand the error characteristics of learned solutions for dynamical systems. 


\paragraph{Neural network-based integration:} With \ac{NN}-based integration methods we \textit{learn} an approximate solution function to a differential equation or a system of differential equations. This learned solution offers the upside of fast and explicit evaluations while providing sufficiently accurate results over large time steps, given appropriate training. These benefits come at the expense of the need to train the model in advance. As suggested in \cite{lagaris_artificial_1998} the training procedure can be set up as a collocation method. The NN constitutes the candidate function and is fit to match the differential equations at collocation points. This methodology has been revived in \cite{raissi_physics-informed_2018} under the term \ac{PINN}.
A related approach, that can be seen as a generalisation of PINNs, is termed operator learning \citep{lu_learning_2021,kovachki_neural_2023}; we will subsequently use NN or PINN to refer to these kind of approaches. As these methods do not adhere to error characteristics by construction---unlike \ac{RK} schemes---understanding and estimating the prediction errors is an open research question.


\paragraph{Approaches for analysing the error of PINNs:} 
The trivial approach consists in comparing the \ac{PINN}'s prediction with a highly accurate integration scheme that can be regarded as ground truth. However, as this assessment might involve significant computational cost, our interest lies on approaches for estimating the error characteristics without knowing the ground truth. For the simplest version of \acp{PINN}, we obtain a local approximation error $\mathcal{O}(1)$, meaning that the error is independent of the time step size $h$. \cite{lagaris_artificial_1998} proposed to change the \ac{NN}'s architecture to ensure numerical consistency for small $h$; reducing the time step size should reduce the error. However, achieving high orders $p$ would require many additional function evaluations which reduces the attractiveness of \acp{PINN} from the standpoint of speed. Furthermore, these changes do not necessarily help with controlling the error for large time steps. This is undesirable as \acp{PINN} become an attractive alternative to \ac{RK} schemes only for large time steps. 

The following approaches focus on providing guaranteed error bounds to certify a \ac{PINN}. \cite{hillebrecht_certified_2022} formulate error dynamics between the exact solution and the \ac{PINN} approximation. These dynamics can be bounded using the Lipschitz constant of the dynamical system, but for non-linear systems this might not result in tight error bounds. Several works \citep{yarotsky_error_2017,de_ryck_approximation_2021,de_ryck_generic_2022} aim to provide theoretical upper bounds on the approximation errors of \acp{NN} based on the function class that they describe, however, these results do not provide practical error estimates. \cite{eiras_provably_2023} verify that the residual of the differential equations based on the learned approximation is bounded and they show empirically that this bound is related to the maximum approximation error.


\paragraph{Proposed approach:} In this work, we combine \acs{PINN} with the collocation method of Gauss-Legendre Runge-Kutta schemes, a special form of implicit RK schemes, to estimate the error, similar to embedded RK schemes. To this end, we evaluate the \ac{PINN} at time instances (nodes) defined by the IRK scheme. For a high-order IRK scheme, the function values at the nodes coincide with intermediate values along the trajectory of interest. We then evaluate how closely these predictions match the implicit system of equations defined by the IRK scheme. Based on the mismatch, we can estimate the error at the nodes and hence along the trajectory. As a numerical example, we illustrate the methodology on the logistic equation. A power system generator model with four states serves as a second example to showcase the resulting error estimation on a relevant use case.

\paragraph{Paper structure:}\Cref{sec:dynamical_system_approximation} briefly introduces the notation as well as IRK methods and PINNs as numerical integration methods. \Cref{sec:error_estimate} explains the error estimation procedure. \Cref{sec:results_logistic,sec:results_generator} illustrate the methodology on the example of the logistic equation and the power system generator respectively. \Cref{sec:conclusions} concludes and discusses next steps.

\section{Approximating the trajectory of a dynamical system}\label{sec:dynamical_system_approximation}

We assume an autonomous dynamical system governed by
\begin{align}
    \ddt \xstate(t) = \fupdateof{\xstate(t)}\label{eq:f_update}
\end{align}
with state $\xstate{} \in \Rdim{n}$, time $t \in \Rdim{}$ and the update function $\fupdate: \Rdim{n} \mapsto \Rdim{n}$. Given an initial condition $\xstate_0 = \xstate{}(t_0)$, the trajectory $\xstate(t_0 + h)$ can be obtained by integration of \cref{eq:f_update}
\begin{align}
    \xstate(t_0 + h) = \xstate_0 + \int_{t_0}^{t_0 + h} \fupdateof{\xstate(t)} dt\label{eq:trajectory_definition}.
\end{align}
We refer to $h = t - t_0$ as the time step size. Given that $\fupdate$ is continuous and Lipschitz continuous, the existence and uniqueness of $\xstate(h)$ is ensured. Usually, no analytical form for \cref{eq:trajectory_definition} exists, hence, we revert to constructing an approximation $\xstatehat(h)$ of the trajectory. There is a great variety of established methods, see for example \cite{hairer_solving_1993,iserles_first_2008}, in this work we focus on implicit \ac{RK} methods (\cref{subsec:irk_methods}). In \cref{subsec:PINNs}, we present \acp{PINN} as a \ac{NN}-based integration method.

\subsection{Implicit Runge-Kutta (IRK) methods}\label{subsec:irk_methods}

As described in \cite{iserles_first_2008}, \cref{eq:trajectory_definition} constitutes a quadrature problem and it can be approximated by a weighted sum of the function values at \nstages{} nodes
\begin{align}
    \xstate(t_0 + h) \approx \xstate{}_0 + h \sum_{i=1}^{\nstages} b_i \, \fupdateof{\xstate(t_0 + c_i h)}.
\end{align}
The location of each node is determined by the parameter $c_i$ which usually ranges in $[0, 1]$ and it is associated with the positive weight $b_i$. As the quadrature requires $\xstate(t_0 + c_i h)$ at the nodes, their values are approximated by $\rkstage{i} \approx \xstate(t_0 + c_i h)$ and are subsequently referred to as the $i$-th stage of the \ac{RK} scheme. By constructing a polynomial of order \nstages{} that passes through the node values \rkstage{i}, a set of conditions can be derived---we construct a collocation method. The conditions are described by the following system of implicit equations
\begin{align}\label{eq:irk_system}
    \rkstage{i} = \xstate_0 + h \sum_{j=1}^{\nstages} a_{ij} \fupdateof{\rkstage{j}}.
\end{align}
By solving for \rkstage{i}, we can then calculate 
\begin{align}
    \xstatehat(t_0 + h) = \xstate{}_0 + h \sum_{i=1}^{\nstages} b_i \,\fupdateof{\rkstage{i}}.
\end{align}
The number of stages \nstages{} and the choice of $a_{ij}, b_i, c_i$ determine the accuracy of the scheme. It can be shown, that the choice of Gauss-Legendre polynomials for $b_i, c_i$ leads to a scheme of numerical order $2\nstages{}$. This implies that the approximation error scales according to $\mathcal{O}\left(h^{2\nstages{}}\right)$. This relationship allows IRK schemes to achieve highly accurate approximations for large time step sizes $h$ and moderate numbers of stages $s$. The approximation error $\errorcollocation{i} \in \Rdim{n}$
\begin{align}\label{eq:collocation_error}
    \errorcollocation{i} := \rkstage{i} - \xstate(t_0 + c_i h) 
\end{align}
of the stages that result from solving the collocation scheme scales as $\mathcal{O}\left(h^\nstages\right)$ \citep{butcher_practical_2009}. We will use this relation in the error estimation in \cref{sec:error_estimate} as it provides insightful information at intermediate points along the trajectory.


\subsection{Neural-Network-based approximation}\label{subsec:PINNs}

In the previous section, the candidate solution for the collocation problem is a polynomial. Instead, we now choose a NN to approximate \xstate{}
\begin{align}\label{eq:NN_approximation}
    \xstatehat(t_0 + h) = \xstate{}_0 + h \, \NN{}_{\theta}(h).
\end{align}
The modifications of multiplying $\NN{}_{\theta}(h)$ with $h$ and the addition of the initial condition $\xstate{}_0$ can already be found in \cite{lagaris_artificial_1998}. They ensure that the initial condition is satisfied for $h=0$ and the local approximation error will scale as $\mathcal{O}\left(h\right)$ if $\NN{}_{\theta}(h)$ is bounded.

In this work, we use a simple feed-forward neural network $\NN{}_{\theta}: \Rdim{d_0} \rightarrow \Rdim{d_L}$ with $L$ layers of size $d_l$. It describes a concatenation of affine transformations $\mathcal{A}_l : \Rdim{d_{l-1}} \rightarrow \Rdim{d_l}: z \mapsto W_l z + b_l$ with non-linear element-wise activation functions $\sigma: \Rdim{} \rightarrow \Rdim{}$
\begin{align*}
    \NN{}_{\theta}(h) =\mathcal{A}_L \circ \sigma \circ \mathcal{A}_{L-1} \circ \dots \circ \sigma \circ \mathcal{A}_{1}(h).
\end{align*}
The weight matrices $W_l$ and bias vectors $b_l$ of the $L$ layers constitute the trainable parameters $\theta$ and for the present case, the time step size forms the input $(d_0 = 1)$ and the $n$ states of the system form the output $(d_L = n)$. The training problem to determine $\theta$ becomes
\begin{align}\label{eq:loss_pinn}
    \min_{\theta} \frac{1}{N} \sum_{j=1}^{N} \norm{\ddt \xstatehat\left(t_0 + h^{(j)}\right) - \fupdateof{\xstatehat\left(t_0 + h^{(j)}\right)}}{2}^2
\end{align}
as introduced in \cite{lagaris_artificial_1998} and termed \acp{PINN} by \cite{raissi_physics-informed_2018}. This can be seen as a collocation method for finding $\theta$ by evaluating the norm in \cref{eq:loss_pinn} at $N$ collocation points at time step sizes $h^{(j)}$. The temporal derivative $\ddt\xstatehat(t_0 + h^{(j)})$ can be computed using automatic differentiation (AD) \citep{baydin_automatic_2018}, hence, this training procedure does not require simulated data points. We solve the optimisation problem in \cref{eq:loss_pinn} with a gradient descent method, namely the limited-memory Broyden–Fletcher–Goldfarb–Shanno algorithm (L-BFGS) \citep{liu_limited_1989}.

\section{Error estimates of PINN predictions with IRK methods}\label{sec:error_estimate}

The following error estimation approach is closely linked to the works in \cite{hairer_solving_1996,de_swart_construction_1997,butcher_practical_2009}, however, we do not focus on the error estimation at the end of the time step $|\xstatehat(t_0 + h) - \xstate(t_0 + h)|$. Instead, we aim for \nstages{} error estimates at the \nstages{} locations of the RK stages $t_0 + c_i h$ which are spread across the interval $[t_0, t_0 + h]$ as $c_i \in [0, 1]$. 

To this end, we evaluate the approximation function $\xstatehat$ defined by the PINN at the node locations $t_0+ c_i h$ that correspond to an IRK method of our choice. These predictions act as approximations 
\begin{align}\label{eq:stage_approximation}
    \rkstagehat{i}=\xstatehat{}\left(t_0+ c_i h\right)
\end{align}
of the IRK stages \rkstage{i} and will, generally speaking, lead to a mismatch $\rkstagedefect{i} \in \Rdim{n}$
\begin{align}\label{eq:stage_error}
    \rkstagedefect{i} := \rkstagehat{i} - \rkstage{i},
\end{align} 
that we will call the \textit{stage error}. Using \cref{eq:collocation_error,eq:stage_approximation,eq:stage_error} and $\errorcollocation{i} = \mathcal{O}\left(h^{\nstages}\right)$, we can formulate the desired error estimate for the NN-based function approximation
\begin{align}\label{eq:error_approximation_NN}
    \xstatehat(t_0 + c_i h) - \xstate(t_0 + c_i h) = \rkstagehat{i} - (\rkstage{i} + \errorcollocation{i}) =  \rkstagedefect{i} + \mathcal{O}\left(h^{\nstages}\right)
\end{align}
The subsequent paragraph describes the estimation of \rkstagedefect{i} without explicitly solving for \rkstage{i}.



\paragraph{Estimation of \rkstagedefect{i}:}
Due to the stage error \rkstagedefect{i}, the corresponding system of equations in \cref{eq:irk_system} evaluated with \rkstagehat{i} will also not be satisfied; a residual $\rkresidual{i} \in \Rdim{n}$ for the system of equations for the $i$-th stage remains
\begin{align}
    \rkresidual{i} := \rkstagehat{i} - \xstate_0 - h \sum_{j=1}^{\nstages} a_{ij} \fupdateof{\rkstagehat{j}}.\label{eq:irk_nn_residual}
\end{align}
We begin by replacing $\rkstagehat{i} = \rkstage{i} + \rkstagedefect{i}$ and expanding $\fupdateof{\rkstage{i} + \rkstagedefect{i}}$ in a Taylor series expansion
\begin{align}\label{eq:error_expansion}
\begin{split}
    \rkresidual{i} &=  \rkstage{i} + \rkstagedefect{i} - \left(\xstate(t_0) + h \sum_{j=1}^{\nstages} a_{ij} \fupdateof{\rkstage{j} + \rkstagedefect{j}}\right)\\
    &= \rkstage{i} + \rkstagedefect{i} - \left(\xstate(t_0) + h \sum_{j=1}^{\nstages} a_{ij} \left[\fupdateof{\rkstage{j}} + \jacobianfx{\rkstage{j}} \rkstagedefect{j} + \mathcal{O}\left({\rkstagedefect{j}}^2\right)\right]  \right)\\
    &= \rkstagedefect{i} -  h \sum_{j=1}^{\nstages} a_{ij}  \jacobianfx{\rkstage{j}} \rkstagedefect{j} + \mathcal{O}\left({\rkstagedefect{j}}^2\right)
\end{split}
\end{align}
The last simplification follows from the definition of the IRK scheme evaluated on \rkstage{i} as it must equal $0$. As \cref{eq:error_expansion} depends on \rkstage{j}, we perform a Taylor series expansion of \jacobianfx{\rkstage{j}} using $\rkstage{j} = \rkstagehat{j} - \rkstagedefect{j}$. The relation between \rkresidual{i} and \rkstagedefect{i} becomes
\begin{align}
    \rkresidual{i} &= \rkstagedefect{i} -  h \sum_{j=1}^{\nstages} a_{ij}  \jacobianfx{\rkstagehat{j}} \rkstagedefect{j} + \mathcal{O}\left({\rkstagedefect{j}}^2\right).
\end{align}
By collecting $\rkstagedefectall = \begin{bmatrix}
    \rkstagedefect{1} & \dots & \rkstagedefect{\nstages}
\end{bmatrix} \in \Rdim{n\nstages}$ and $\rkresidualall = \begin{bmatrix}
    \rkresidual{1} & \dots & \rkresidual{\nstages}
\end{bmatrix} \in \Rdim{n\nstages}$ we obtain
\begin{align}
    \rkresidualall &= \left( I - h J\right) \rkstagedefectall  + \mathcal{O}\left({\rkstagedefectall}^2\right)\end{align}
with the identity matrix $I\in \Rdim{n\nstages \times n\nstages}$ and the Jacobian-like matrix $J\in \Rdim{n\nstages \times n\nstages}$ that is constructed based on the Jacobians $\jacobianfx{\rkstagehat{i}} \in  \Rdim{n\times n}$ and the weights of the IRK scheme $a_{ij}$
\begin{align*}
    J = \begin{bmatrix} a_{11}  \jacobianfx{\rkstagehat{1}} & \dots &  a_{1\nstages}  \jacobianfx{\rkstagehat{\nstages}} \\ \vdots & \ddots & \vdots \\ a_{\nstages 1} \jacobianfx{\rkstagehat{1}} & \dots &  a_{\nstages\nstages} \jacobianfx{\rkstagehat{\nstages}} \end{bmatrix}.
\end{align*}
Neglecting the higher order terms of $\rkstagedefectall$ and inverting $(I - hJ)$, we obtain
\begin{align}\label{eq:error_estimate_delta}
    \rkstagedefectall{} &= \left( I - h J\right)^{-1} \rkresidualall{}
\end{align}
which can then be used in \cref{eq:error_approximation_NN} to estimate the prediction error of the NN.

\section{Logistic equation - illustration with an analytical solution}\label{sec:results_logistic}

This section demonstrates the methodology on the logistic equation
\begin{align*}
    \ddt x = x ( 1 - x)\label{eq:logistic}
\end{align*}
with the analytical solution for an initial conditions $x(t_0) > 0$ 
\begin{align*}
    x(t_0 + h) = \frac{e^h}{c_0 +  e^h}; \qquad c_0 = \frac{1}{x(t_0)} - 1.
\end{align*}
We use the simple but nonlinear logistic equation due to the ease of visualisation and the existence of an analytical solution. The arguments can be equally made for dynamical systems with multiple states as we demonstrate later on. To build intuition, we show in \cref{subsec:results_predictors} the prediction accuracy of different integration schemes and in \cref{subsec:results_stages} we illustrate the interpretation of the IRK stages as approximate intermediate values $\xstate(t_0 + c_i h)$. \Cref{subsec:results_estimation} then presents the error estimation. 

\subsection{Approximating a trajectory of the logistic equation}\label{subsec:results_predictors}

\begin{figure}[ht]
    \centering
    \subfigure[\small Trace of predictions]{
      \includegraphics[height=4.5cm]{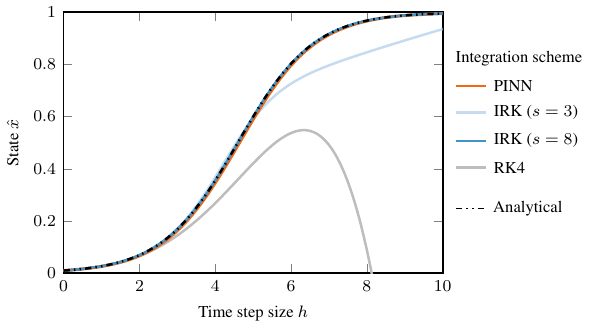}\label{fig:trajectory_trace}}
    \hfill
    \subfigure[\small Prediction errors]{
      \includegraphics[height=4.5cm]{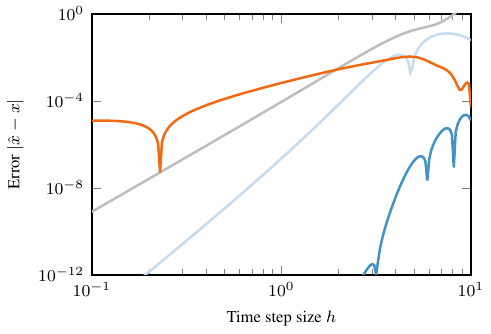}\label{fig:error_order}}
    \caption{Predictions from different numerical integration schemes for a time step of size $h$.}
    \label{fig:approximation_scheme}
\end{figure}

We apply different integration schemes to the logistic equation, starting with an initial condition $x_0 = 0.01$. \Cref{fig:trajectory_trace} shows the trace of the predictions when a single time step of size $h$ is applied. We first want to note that---despite the simplicity of the trajectory---an explicit RK scheme such as the well-known fourth-order RK4 scheme fails to predict large time steps. Hence, we revert to IRK schemes for predictions over long time steps. With three stages ($s=3$), we can still observe an error, for eight stages ($s=8$), the predicted values are indistinguishable from the analytical solution. Lastly, we plot the prediction of a trained PINN and it also closely matches the analytical solution over the entire range of the tested time step sizes.

Plotting the approximation error $|\hat{x} - x|$ against the time step size in \cref{fig:error_order} provides further insights. By showing both axes on a logarithmic scale the order $p$ of the different schemes become visible. For the PINN, such a convergence properties is not observed, in fact, the error is significantly higher for $h =5$ than for $h=10$. This example illustrates, that while PINNs can predict large time steps accurately, reducing the time step size can lead to larger errors. This creates an error characteristic that is more difficult to analyse and control than for a RK scheme.

\subsection{Visualising the stages of an IRK scheme}\label{subsec:results_stages}

\begin{figure}[th]
    \centering
    \subfigure[\small Trace of stage values $k^{(i)}$]{
      \includegraphics[height=4.4cm]{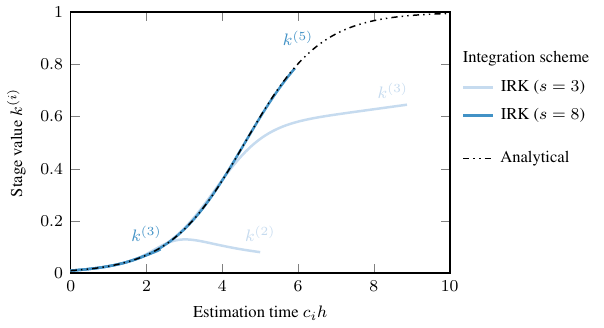}\label{fig:stage_trace}}
    \hfill
    \subfigure[\small Stage approximation errors]{
      \includegraphics[height=4.4cm]{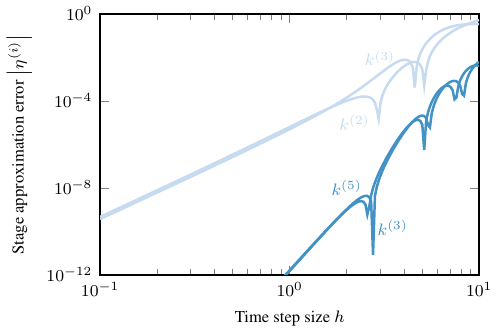}\label{fig:stage_error_order}}
    \caption{Analysis of the prediction accuracy of IRK stages for a time step of size $h$.}
    \label{fig:stage_values}
\end{figure}

We now perform a similar analysis for the stage value $k^{(i)}$ of the IRK schemes. We plot their values against the time $c_i h$ that corresponds to the $i$-th node. For $s=3$, we can observe that the stages $k^{(2)}$ and $k^{(3)}$ in \cref{fig:stage_trace} follow the analytical solution well at first, but then deviate. For the higher order IRK scheme $(\nstages = 8)$ the stages $k^{(3)}$ and $k^{(5)}$ coincide closely with the analytical solution, that is the true trajectory. 
\Cref{fig:stage_error_order} underlines this observation as we plot the stage approximation error $\eta^{(i)}$ against the time step size $h$. This error follows the order $\mathcal{O}\left(h^{\nstages}\right)$, which illustrates that interpreting the stage values $k^{(i)}$ as approximations of the trajectory itself at intermediate points becomes very precise for high order schemes, even for large time step sizes $h$. The proposed error estimate exploits this fact.

\subsection{Estimating the stage error}\label{subsec:results_estimation}
\begin{figure}[th]
    \centering
    \includegraphics[]{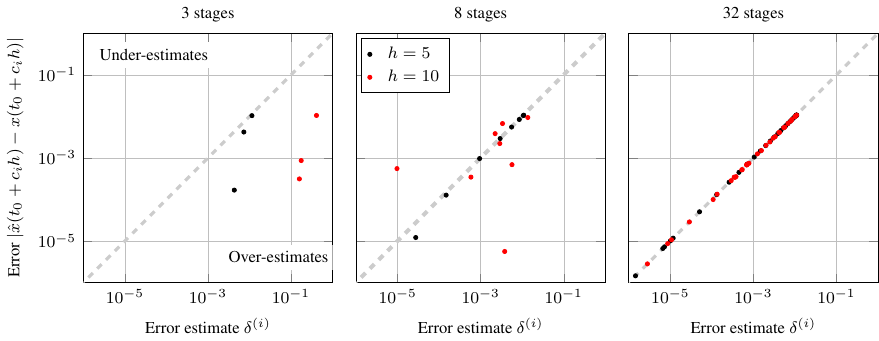}
    \caption{Correlation between the estimated stage error $\delta^{(i)}$ and the corresponding prediction error $|\hat{x}(t_0 + c_i h) - x(t_0 + c_i h)|$ for IRK schemes with 3, 8, and 32 stages. The black dots represent a time step size of $h=5$ and the red dots a time step size of $h=10$.}
    \label{fig:error_correlation}
\end{figure}

With this intuition in mind, we now apply the methodology presented in \cref{sec:error_estimate}. The question we want to ask is how well the estimated error \rkstagedefect{i} correlates with the prediction error $\xstatehat(t_0 + c_i h) - \xstate(t_0 + c_i h)$. A strong correlation improves the usefulness of the estimate as it is more reliable. As the analytical solution exists, we compute the two errors for a time step of $h=5$ (black dots) and $h=10$ (red dots) and show the result in \cref{fig:error_correlation}.
 The subplots show the correlation for different number of stages \nstages{}. The dashed grey line indicates identity between the estimated and true predicted error. It is clearly visible that the larger \nstages{} and the smaller time step size $h$ improve the correlation which can be expected as the effect of the stage error $\eta^{(i)}$ reduces (see \cref{subsec:results_stages}).

\begin{figure}[!bh]
    \centering
    \includegraphics[]{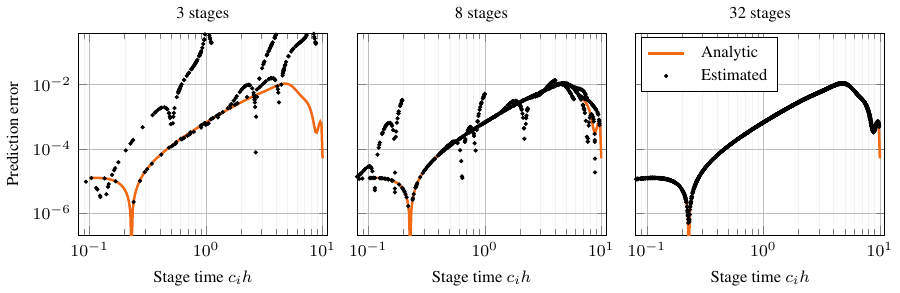}
    \caption{Consistency of stage prediction for different orders of the IRK schemes. The points indicate the error estimates $\delta^{(i)}$ for a range of values of $h$. The exact error $|\xstatehat - \xstate|$ is shown in orange. For low order IRK schemes, the error estimates for a given time $c_i h$ clearly contradict each other. For a high-order IRK scheme all error estimates form a consistent estimate that is in line with the exact error characteristics.}
    \label{fig:stage_consistency}
\end{figure}

To identify the number of stages that is required for a reliable estimate, we plot the estimated error $\delta^{(i)}$ against the associated time value $c_i h$ for different values of $h$, shown in \cref{fig:stage_consistency}. We observe that for an IRK scheme with few stages (left plot), nearby values of $c_i h$ have orders of magnitude different error estimates $\delta^{(i)}$. By increasing the number of stages, the error estimates become more consistent and approach the error characteristic based on the analytical solution.
If the analytical solution is unknown, checking the consistency of the error estimate along a trajectory can be used for validating the estimation accuracy.


\section{Error estimation for a power system generator model}\label{sec:results_generator}

As motivated in \cref{sec:introduction}, NN-based approximations might be able to accelerate the simulation of dynamical phenomena in power systems. The PINNSim approach \citep{stiasny_pinnsim_2023} relies on learning the dynamics of single components, such as generators, and then resolving their interactions. Being able to estimate errors of the learned dynamics will be an important step for future development of such approaches. As an example for the dynamics of a component, we use a two-axis generator model which is modelled by a state vector $\xstate \in \Rdim{4}$ governed by
\begin{align*}
    \ddt \begin{bmatrix}
        E'_q \\ E'_d \\ \delta \\ \Delta \omega
    \end{bmatrix} &= \begin{bmatrix}
        \frac{1}{T'_{do}}\left(-E'_q - \left(X_d - X'_d\right) I_d + E_{fd}\right)\\
        \frac{1}{T'_{qo}}\left(-E'_d + \left(X_q - X'_q\right) I_q\right)\\
        \omega_s \Delta \omega\\
        \frac{1}{2H}\left(P_m - E'_d I_d - E'_q I_q - \left(X'_q - X'_d\right) I_d I_q - D \Delta \omega\right)
    \end{bmatrix}\\
    \begin{bmatrix}
        I_d\\I_q
    \end{bmatrix} &= \begin{bmatrix}
        R_s & -X_q\\X_d & R_s
    \end{bmatrix}^{-1} \begin{bmatrix}
        E'_d - V \sin{(\delta  - \theta)}\\
        E'_q - V \cos{(\delta  - \theta)}
    \end{bmatrix}.
\end{align*}
We refer to \cite{sauer_power_1998} for a detailed description of the modelling and the used parameters. The following results use the parameters associated with generator 2 in \citet[p.~167]{sauer_power_1998}.

\begin{figure}[th]
    \centering
    \subfigure[Estimated error]{
      \includegraphics[height=4.4cm]{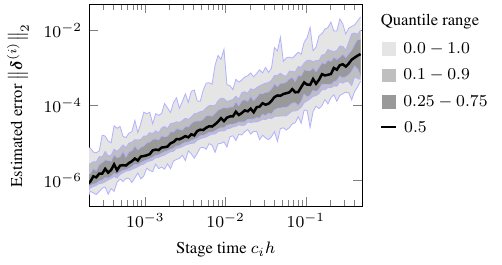}}
    \hfill
    \subfigure[Error evaluated with a solver]{
      \includegraphics[height=4.4cm]{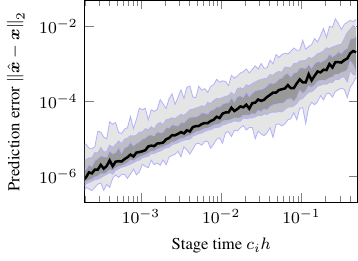}}
    \hfill
    \caption{Error distribution of the PINN predictions with respect to the time step size $h$ for a two-axis generator model. We compare the estimated errors \norm{\rkstagedefect{i}}{2} to the prediction error evaluated with a highly accurate solver. As the error depends also on the initial condition, the grey shadings correspond to the errors for 100\%, 80\%, and 50\% of these trajectories.}
    \label{fig:generator_errors}
\end{figure}

In \cref{sec:results_logistic}, we learned a single trajectory. Here, we aim to predict a collection of trajectories starting from initial conditions $\xstate_0 \in \mathcal{X}_0$ within the set $\mathcal{X}_0$, hence, we adjust \cref{eq:NN_approximation} to consider $\xstate_0$ as an additional input to the PINN 
\begin{align}
    \xstatehat{}(t_0 + h, \xstate_0) = \xstate_0 + h \, \NN{}_{\theta}(h, \xstate_0).
\end{align}
As a consequence, the approximation error will depend on $h$ and $\xstate_0$. Therefore, we sample initial conditions from $\mathcal{X}_0$ and apply the proposed error estimation technique for each $\xstate_0$. As a result we obtain $s$ error estimates along the trajectory of every tested $\xstate_0$. To report the error, we compute a norm over the prediction error $\xstatehat{} - \xstate{}$ (evaluated at $c_i h$) and the estimated error $\rkstagedefect{i}$
\begin{align*}
    \norm{\xstatehat{} - \xstate{}}{2} &= \sqrt{\sum_j^n \xi_j^2 \left( \hat{x}_j - x_j \right)^2}\\
    \norm{\rkstagedefect{i}}{2} &= \sqrt{\sum_j^n \xi_j^2 {\delta_j^{(i)}}^2}.
\end{align*}
We include a scaling factor $\xi_j$ in the norm so that all state variations have similar magnitude, see \citet{stiasny_physics-informed_2023-1} for more details. This factor was also employed when evaluating \cref{eq:loss_pinn} during training the PINN model (3 layers with 64 neurons each). As for \cref{fig:stage_consistency}, we collect the estimated errors at all $s$ stages. \Cref{fig:generator_errors} shows the resulting error estimates $\norm{\rkstagedefect{i}}{2}$ and the prediction errors evaluated with a differential equation solver with tolerance settings of $10^{-12}$, as no analytical solution is available. To represent the dependency of the prediction error on the different initial conditions, we compute quantiles over small intervals of $c_i h$ which corresponds to the location of the stages. 
\Cref{fig:generator_errors} uses $s=8$ and shows a great agreement of the estimated error with the true prediction error apart from a few instances where the proposed error estimation scheme over-estimates the maximum error, for example at $c_i h = 0.01$ and $c_i h = 0.05$. By increasing the number of stages, the estimation quality can be further improved.
\section{Conclusions}\label{sec:conclusions}

The estimation of prediction errors poses a major challenge for \acfp{PINN}. This work introduces a simulation-free and practical error estimation approach, which can remove the barrier for incorporating \acp{PINN} in established numerical solvers. 
Our methodology only requires the evaluation of an implicit Runge-Kutta (IRK) scheme (whose coefficients are usually pre-computed) and a matrix inversion, which leads to a simple and easy-to-implement estimation method, see \cref{eq:error_estimate_delta}. We demonstrate that the proposed approach can accurately estimate the prediction errors along a trajectory and is also applicable for dynamical systems with multiple states. We also show that it is straightforward to assess the quality of this estimate by analysing the consistency of the error estimates that stem from different time step sizes $h$.

\paragraph{Future work:} From a theoretical perspective, we see the need to further investigate the influencing factors on the estimation accuracy. That is, first, the analysis of the impact of higher-order terms of $\rkstagedefect{i}$ which could become significant if the \ac{PINN}'s prediction is not very accurate. Furthermore, the impact of system characteristics such as stiffness and oscillatory behaviour on the required number of stages $s$ should be investigated. 

With respect to practical applications, the proposed estimation methodology can be used in monitoring the training progress, assessing the learned model, or providing estimates during prediction tasks. We plan to explore these use cases to develop heuristics for this method. Furthermore, the methodology could inform an additional term in the loss function similar to \cite{raissi_physics-informed_2018,stiasny_physics-informed_2023-1}. Lastly, the error estimates could also be employed at the evaluation stage to correct the \ac{PINN}'s prediction. 

\newpage
\acks{This work was supported by the ERC Project VeriPhIED, funded by the European Research Council, Grant Agreement No: 949899. The authors would like to thank Samuel Chevalier, Ignasi Ventura Nadal, and Petros Ellinas for their feedback on the manuscript.}
\bibliography{references_manual}

\end{document}